\documentclass[a4paper,man,floatsintext,longtable,noextraspace,12pt]{apa6}

\usepackage[english]{babel}
\usepackage[utf8x]{inputenc}
\usepackage{amsmath}
\usepackage{graphicx}
\usepackage[colorinlistoftodos]{todonotes}
\usepackage{hyperref}

\usepackage{booktabs}
\usepackage{longtable}
\usepackage{array}
\usepackage{multirow}
\usepackage{wrapfig}
\usepackage{float}
\usepackage{colortbl}
\usepackage{pdflscape}
\usepackage{tabu}
\usepackage{threeparttable}
\usepackage{threeparttablex}
\usepackage[normalem]{ulem}
\usepackage{makecell}
\usepackage{xcolor}
\NewDocumentCommand\citeproctext{}{}

\makeatletter
 \let\@cite@ofmt\@firstofone
 \def\@biblabel#1{}
 \def\@cite#1#2{{#1\if@tempswa , #2\fi}}
\makeatother
\newlength{\cslhangindent}
\newlength{\csllabelwidth}
\newenvironment{CSLReferences}[2] % #1 hanging-indent, #2 entry-spacing
  {\begin{list}{}{%
   \setlength{\itemindent}{0pt}
   \setlength{\leftmargin}{0pt}
   \setlength{\parsep}{0pt}
   \ifodd #1
    \setlength{\leftmargin}{\cslhangindent}
    \setlength{\itemindent}{-1\cslhangindent}
   \fi
   \setlength{\itemsep}{#2\baselineskip}}}
  {\end{list}}
\usepackage{calc}

\providecommand{\tightlist}{%
  \setlength{\itemsep}{0pt}\setlength{\parskip}{0pt}}

\title{\textbf{How open are hybrid journals included in transformative agreements?}}
\shorttitle{Hybrid OA}
\author{Najko Jahn}
\affiliation{Göttingen State and University Library, University of Göttingen\\
Platz der Göttinger Sieben 1, 37073 Göttingen, Germany\\
najko.jahn@sub.uni-goettingen.de
}

\authornote{ORCID: https://orcid.org/0000-0001-5105-1463}

\abstract{The ongoing controversy surrounding transformative agreements, which aim to transition subscription-based journal publishing to full open access, highlights the need for large-scale studies assessing the impact of these agreements on hybrid open access. By combining multiple open data sources, including cOAlition S Journal Checker, Crossref, and OpenAlex, this study presents a novel approach that analyses over 700 agreements. Results suggest a strong growth in open access, from 4.3\% in 2018 to 15\% in 2022. Over five years, 11,189 hybrid journals provided open access to 742,369 out of 8,146,958 articles (9.1\%). Authors who could make use of transformative agreements contributed 328,957 open access articles (44\%) during this period, reaching a peak in 2022 with 143,615 out of 249,511 open access articles (58\%). While this trend was predominantly driven by the three commercial publishers Elsevier, Springer Nature, and Wiley, open access uptake varied substantially across journals, publishers, disciplines, and countries. Particularly, the OECD and BRICS areas revealed different publication trends. In conclusion, this study suggests that current levels of implementation of transformative agreements is insufficient to bring about a large-scale transition to full open access.}


\begin{document}
\maketitle

% QSS wants numbered sections
\setcounter{secnumdepth}{2}

\section{Introduction}\label{introduction}

For over two decades, hybrid open access journal publishing, which makes
some articles openly available while others remain behind a paywall, has
been discussed as a means for transitioning the subscription system to
full open access (Prosser, 2003). The idea was that when journals
increasingly publish open access articles, they could reduce revenues
from subscriptions, while libraries and funders could change their
funding models and shift expenditures from subscription to open access.
However, initial approaches, mainly based on publication fees, also
called article processing charges (APCs), did not contribute
substantially to a large open access uptake. In 2009, the publisher
Springer reported that 1\% of its articles in hybrid journals were open
access (Dallmeier-Tiessen et al., 2010). Other studies have also
recorded low uptake. In 2011, only 1-2\% of articles were open access
(Björk, 2012), increasing to around 4\% between 2011 and 2013 (Laakso \&
Björk, 2016).

With the introduction of central funding mechanisms for publication fees
in some European countries since 2012, an substantial increase in hybrid
open access has been observed (Björk, 2017; Huang et al., 2020; Jubb et
al., 2017; Piwowar et al., 2018). For example, studying university
output, Robinson-Garcia et al. (2020) estimated a median uptake of 7.1\%
in the period 2014-2017. In particular, British (17\%), Austrian (15\%)
and Dutch (13\%) universities stood out. However, this shift in funding
policy towards hybrid open access also added to the overall cost of
publishing, which includes subscription spending and the administrative
efforts required to handle payments (Pinfield et al., 2016). Moreover,
large commercial publishers, which already dominated the publishing
market (Larivière et al., 2015), disproportionately benefited from
hybrid open access funding in comparison to full open access publishers
(Butler et al., 2023; Jahn \& Tullney, 2016; Shu \& Larivière, 2023).

As a consequence, libraries and their consortia began to develop
licensing strategies aimed at avoiding such `double dipping' scenarios,
in which well-established commercial publishers gain twice from reading
and open access publishing fees (Mittermaier, 2015), as well as to
increase publisher-provided immediate open access (Björk \& Solomon,
2014; Schimmer et al., 2015). These considerations resulted in
transformative agreements\footnote{In this paper I use the term
  ``transformative agreement'', addressing also offsetting,
  read-and-publish or publish-and-read deals, and other variants
  (Borrego et al., 2021; Hinchliffe, 2019). Although the term is
  critised as misleading and not useful to describe the different types
  of open access agreements between library consortia and commercial
  publishers (Babini et al., 2022), it is widely used in policy
  discussions and in the research literature.}, which cover a broad
range of contracts between library consortia and publishers from the
mid-2010s onwards, where institutional spending for subscriptions and
open access publishing are considered together (Borrego et al., 2021;
Hinchliffe, 2019). Transformative agreements seek to control costs while
allowing a transitional phase for publishing more open access articles.
Similar to big deals, transformative agreements mainly bundle hybrid and
subscription-only journals from commercial publishers, but aim at a
higher degree of transparency than previous big deals, where contracts
including payments were confidential (Bergstrom et al., 2014).

The introduction of transformative agreements aligns with funding policy
changes, such as the decision made by cOAlition S, a consortium of
funders and research organisations including the European Commission, to
no longer provide financial support for individual publication fees when
publishing in hybrid journals. According to its Plan S launched in 2018,
cOAlition S members only accept hybrid open access through
transformative agreements ``during a transition period that should be as
short as possible'' (Schiltz, 2018). Specifically, they agreed to
support hybrid open access only through transformative agreements from
2021, until the end of 2024. Notably, despite not being part of
cOAlition S, the German Research Foundation (DFG), has also extended its
financial support for hybrid open access through transformative
agreements (Mittermaier, 2021). Previously, the DFG only provided
funding for fully open access journals (Jahn \& Tullney, 2016).

By the end of 2023, many transformative agreements had been implemented,
but the interim outcomes were mixed. The ESAC Transformative Agreement
Registry\footnote{\url{https://esac-initiative.org/about/transformative-agreements/agreement-registry/}},
the largest resource for library consortia to voluntarily and publicly
share their agreements, recorded more than 800 transformative
agreements. These agreements resulted in the publication of up to
900.000 open access articles published in both fully open access and
hybrid journals, according to the accompanying ESAC Market
Watch\footnote{\url{https://esac-initiative.org/market-watch/}}. Library
consortia reported increased open access volume, streamlined payment and
monitoring procedures, as well as extensive utilisation of open access
options by the researchers they serve (Marques \& Stone, 2020; Parmhed
\& Säll, 2023; Pinhasi et al., 2020). The ongoing standardisation of
transformative agreements has contributed to improved transparency in
terms of contracts and publisher-provided article metadata (Marques et
al., 2019; Pinhasi et al., 2021). However, with the growing trend toward
transformative agreements, continued reliance on big deals is perceived
as problematic, because it perpetuates market concentration (Butler et
al., 2023; Shu \& Larivière, 2023). Whether transformative agreements
lead to reduced pricing remains uncertain (Borrego, 2023) and a
substantial transition of hybrid journals towards full open access could
not be observed (Matthias et al., 2019; Momeni et al., 2021). The focus
on large commercial publishers might also increase inequality
(Ross-Hellauer et al., 2022), because transformative agreements' focus
on pay to publish mainly targets institutions from high-income
countries, furthering a questionable journal prestige culture (Babini et
al., 2022). Besides, an editorial-board resignation raised concerns that
transformative agreements might encourage publishers to maximize journal
publication volume ``without regard to quality'' (Rasmussen, 2023).

The controversies surrounding hybrid open access and transformative
agreements have led to varying policy conclusions. For instance, the
British Joint Information Systems Committee (JISC) evaluated its open
access strategy, which included transformative agreements (Brayman et
al., 2024). The evaluation revealed that while these agreements had a
significant impact on the growth of open access in the country, they had
limited effects in facilitating a global shift towards full open access.
As a result, the report advised that the British open access strategy
should be reassessed. After making similar observation, Norwegian and
Swedish universities and their consortia also argued for policy changes
(Holden et al., 2023; Widding, 2024). Furthermore, cOAlition S conclude
its financial support of transformative agreements at the end of 2024,
but continue to view open access resulting from such agreements as
compliant.\footnote{\url{https://www.coalition-s.org/coalition-s-confirms-the-end-of-its-financial-support-for-open-access-publishing-under-transformative-arrangements-after-2024/}}
cOAlition S also removed the majority of hybrid journals from its
Transformative Journal program in 2023 due to publishers' failure to
meet self-defined open access growth targets (Brainard, 2023). By
contrast, the German DEAL consortium announced a five-year
transformative agreement with Elsevier starting in 2024 and also renewed
its contracts with Springer Nature and Wiley until the end of 2028.
Similarly, the Colombia Consortium signed the first transformative
agreements in Latin America (Muñoz-Vélez et al., 2024).

Despite these controversies around transformative agreements as a means
of transitioning journal publishing to full open access, there is
limited evidence available on the uptake of open access in hybrid
journals, and the extent to which this can be attributed to
transformative agreements. Previous studies have focused on specific
countries (Brayman et al., 2024; Haucap et al., 2021; Huang et al.,
2020; Pölönen et al., 2020; Taubert et al., 2023; Waltman \& Lamers,
2022; Wenaas, 2022) or publisher portfolios (Bakker et al., 2024; Fraser
et al., 2023; Jahn et al., 2022; Momeni et al., 2023; Pieper \&
Broschinski, 2018; Schmal, 2024), while large-scale studies relied on
self-reported agreement data (Moskovkin et al., 2022), or used APC
pricing lists (Shu \& Larivière, 2023). In particular, data availability
is a limiting factor when studying the impact of transformative
agreements (Bakker et al., 2024), because bibliometric databases, even
though many allow the retrieval of open access articles in hybrid
journals, do not directly attribute them to specific transformative
agreements. Likewise, article-level open access invoicing and cost data,
which would make it possible to establish a direct link between
transformative agreements and open access publications (Jahn et al.,
2022; Kramer, 2024).

The present study aims to address these limitations by combining
multiple openly available data sources to determine open access uptake
in hybrid journals, while distinguishing between open access through
transformative agreements and other means. With this novel and open
approach, this first large-scale analysis will answer the following
questions:

\begin{itemize}
\tightlist
\item
  What was the number and proportion of open access articles in hybrid
  journals in transformative agreements between 2018 and 2022?
\item
  To what extent did institutions with a transformation agreement
  contribute to open access in hybrid journals?
\end{itemize}

For both research questions, this study will analyse the variability by
publisher, journal subject, and country.

\section{Methods}\label{methods}

\begin{figure}[ht!]

{\centering \includegraphics[width=0.99\linewidth,]{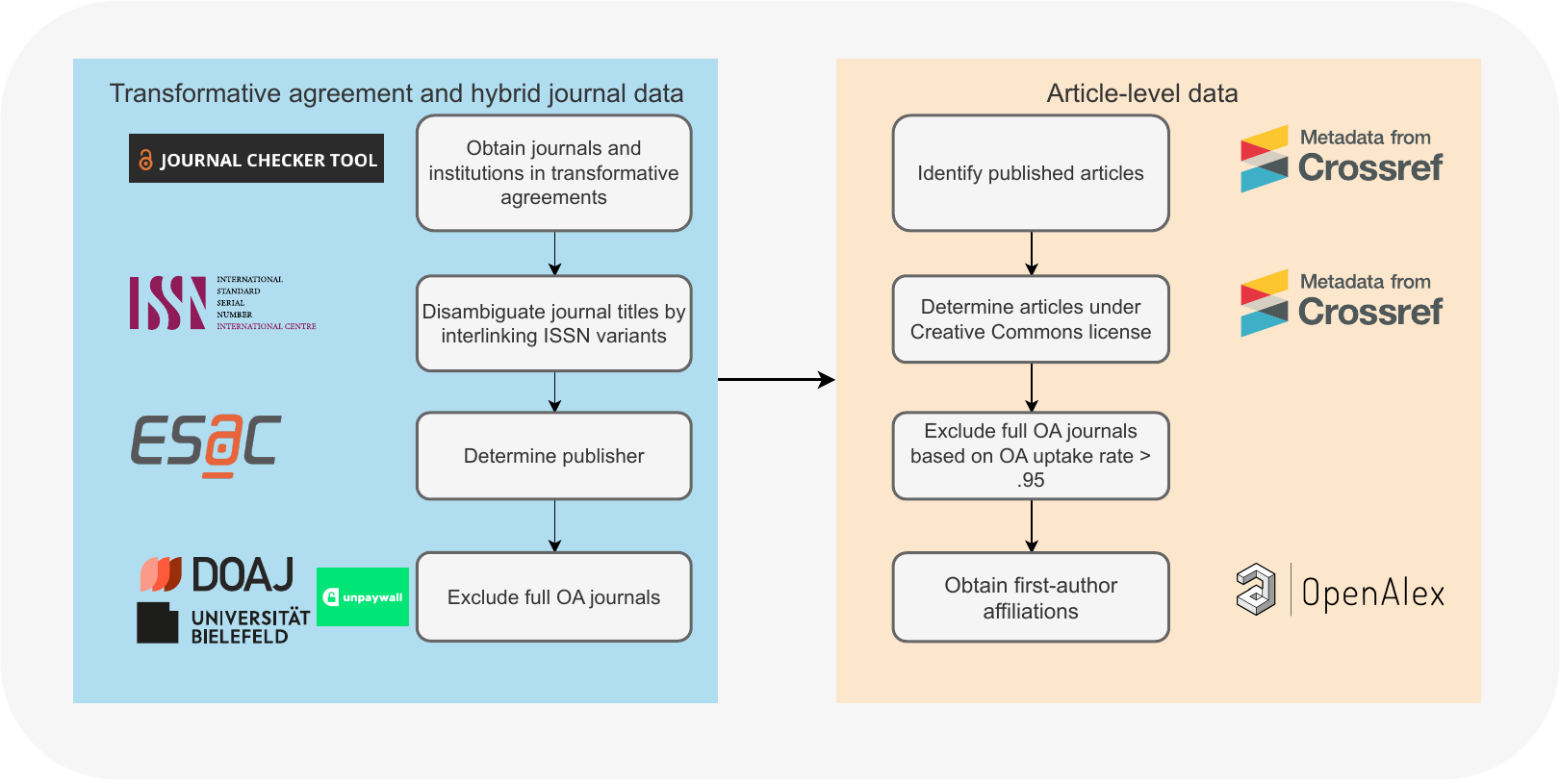} 

}

\caption{Data collection workflow}\label{fig:data_workflow}
\end{figure}

This study combines data from multiple publicly available sources, as
shown in Figure \ref{fig:data_workflow}. Initially, transformative
agreement data retrieved from cOAlition S Journal Checker
Tool\footnote{\url{https://www.coalition-s.org/blog/enabling-accurate-results-within-the-journal-checker-tool/}}
provided information about journal portfolios and participating
institutions. After identifying hybrid journals by excluding fully open
access journals, Crossref (Hendricks et al., 2020) served as the primary
data source for article-level metadata including Creative Commons (CC)
license information to indicate open access availability on publisher
websites. Because of a lack of comprehensive publicly available
invoicing data, open access articles published through transformative
agreements were determined by linking first author affiliations from
OpenAlex (Priem et al., 2022) to eligible institutions according to the
transformative agreement data. In the following, the steps are described
in detail.

\subsection{Transformative agreement and hybrid journal
data}\label{transformative-agreement-and-hybrid-journal-data}

Data gathering started with obtaining journals included in
transformative agreements from the publicly available Transformative
Agreement Data dump\footnote{\url{https://journalcheckertool.org/transformative-agreements/}}
used by the cOAlition S Journal Checker Tool\footnote{\url{https://www.coalition-s.org/blog/enabling-accurate-results-within-the-journal-checker-tool/}},
a voluntary effort based on publicly disclosed contracts in the ESAC
Transformative Agreement Registry\footnote{\url{https://esac-initiative.org/about/transformative-agreements/agreement-registry/}}.
The dump consists of multiple online Google spreadsheets where each data
file represents one agreement listed in the ESAC Transformative
Agreement Registry. From the retrieved spreadsheet files, journals and
institutions involved per agreement were obtained.

It should be noted that although many library consortia see the need to
register their agreements through the ESAC registry, some fail to
disclose all details, including the full-text of contracts (Bakker et
al., 2024; Kramer, 2024). Furthermore, the extent to which ESAC is
comprehensive remains uncertain, potentially limiting the coverage of
transformative agreements in the Journal Checker Tool. Another
limitation of using the Journal Checker Tool and its underlying publicly
available data dump to study the development of transformative
agreements over time is that expired transformative agreements are
constantly removed. To address this, four different snapshots were
safeguarded and combined for this study: self-archived versions from
July 2021, July 2022, and May 2023, as well as the most current dump
downloaded on 11 December 2023. This ensured that transformative
agreements, which ended from 2021 onwards, were included, representing
the majority of transformative agreements. Overall, the four combined
Transformative Agreement Data dumps used in this study contained 729 of
869 agreements listed in the ESAC Transformative Agreement Registry by
December 2023.

The Transformative Agreement Data dumps link agreements to journals
represented by journal names and the ISSN. After mapping ISSN variants
to the corresponding linking ISSN (ISSN-L), as provided by the ISSN
International Centre, journals were associated with publishers according
to the ESAC Transformative Agreement Registry. This reflects that some
portfolios may include imprints. Furthermore, journal subjects according
to the All Science Journal Classification (ASJC) were added from the
Scopus journal source list as of August 2023.

Because transformative agreements can include both fully open access and
hybrid journals, the data were complemented with information about a
journal's open access status using multiple sources: the Directory of
Open Access Journals (DOAJ) downloaded on 12 December 2023\footnote{\url{https://doaj.org/csv}},
OpenAlex (November 2023) and the Bielefeld list of GOLD OA journals
(Bruns et al., 2022). As shown in Figure \ref{fig:method_fig}A,
combining different data sources considerably extended journal matching.
In total, 3,439 full open access journals were excluded based on ISSN
matching. The overlap between the three data sources was 72\%. The Gold
OA journals dataset alone added 176 journals, while the DOAJ comprised
10 fully open access journals not listed in either of the other two
sources. These fully open access journals were launched in 2022.

\subsection{Article and author
metadata}\label{article-and-author-metadata}

After identifying hybrid journals included in transformative agreements,
article metadata were retrieved from the Crossref November 2023 database
snapshot for the five-year period from 2018 to 2022, according to the
issued date, representing the earliest known publication date. Because
Crossref metadata lacked information to distinguish between original
research articles, including reviews, and other types of journal
content, which are often not covered by transformative agreements
(Borrego et al., 2021), only articles published in regular issues
indicated by numeric pagination were included. Furthermore, paratext
recognition was applied to exclude non-scholarly journal content such as
table of contents.

Open access articles in hybrid journals were identified using the
Creative Commons (CC) license information in Crossref metadata, with
consideration of any CC variant. License information relative to the
``accepted manuscript (AM)'' version was not considered. Crossref was
used for open access identification because transformative agreement
workflows generally require publishers to deliver CC license information
to this DOI registration agency (Geschuhn \& Stone, 2017).

Comparing Crossref license coverage with OpenAlex, which re-uses open
access evidence from Unpaywall, a widely used open access discovery
service that also parses journal websites for open content licenses
(Piwowar et al., 2018), highlighted ongoing challenges in identifying
hybrid open access (Butler et al., 2023; Jahn et al., 2022;
Martín-Martín et al., 2018; Zhang et al., 2022). For the purpose of this
study, 742,369 articles under CC license were retrieved using Crossref,
while 950,260 articles were tagged as ``hybrid'' according to the
OpenAlex November 2023 release, which was used throughout this study.
The largest differences concerned articles published between 2018 and
2020. With regard to the publication year 2022, however, Crossref and
OpenAlex open access numbers differ only slightly (249,511 records using
Crossref vs.~255,344 in OpenAlex). Notable differences could be observed
among some publishers that presumably did not provide CC license
information to Crossref, including AIP Publishing, the American
Physiological Society, Emerald, and the Royal Society. Crossref license
metadata was more complete with regard to the articles published by
Wiley and the American Chemical Society. Finally, inconsistent open
access status information in previous OpenAlex versions was observed
(Jahn et al., 2023). After reporting this to OpenAlex, fixing this issue
was still ongoing according to the release notes, which might also
explain part of the discrepancy.

After retrieving the article metadata, the publication volume, including
open access, was calculated for each journal. To improve the
identification of hybrid journals, journals with an open access
proportion above 95\% were excluded. This step allowed removing
additional 134 fully open access journals. Together, these journals
published 8,565 articles between 2018 and 2022.

Affiliation metadata about corresponding authors are crucial for the
planning and evaluation of transformative agreements because they are
considered responsible for arranging open access publication (Borrego et
al., 2021; Geschuhn \& Stone, 2017; Schimmer et al., 2015). For this
study, country and institutional affiliations were retrieved from
OpenAlex. Because the corresponding authorship field was not fully
supported by OpenAlex at the time of analysis, and respective
affiliation data were only available for 54\% of the investigated
articles, this study focused on first authors and their affiliations
instead; approximately 90\% of the articles examined had first author
affiliation metadata in OpenAlex, which is a much larger proportion than
previously reported for the October 2022 snapshot (Zhang et al., 2024).
First authors typically contribute the most to a paper and are often
considered as lead authors (Chinchilla-Rodríguez et al., 2024; Larivière
et al., 2016). In case of lacking data about corresponding authors in
bibliometric databases, related studies also utilised first authors as a
proxy to examine open access payments and transformative agreements
(Haucap et al., 2021; Shu \& Larivière, 2023; Zhang et al., 2022).

To estimate the impact of transformative agreements on hybrid open
access, participating institutions from the Transformative Agreement
Data dump, which the cOAlition S crowd-sourced from agreement documents
and consortia, were matched with the first author affiliations recorded
by OpenAlex using the ROR-ID. Matching also considered the duration of
agreements according to the ESAC registry. In total, 502 agreements were
active between 2018 and 2022. Upon inspection, Transformative Agreement
Data did not comprehensively cover associated institutions, such as
university hospitals or institutes of large research organisations such
as the Max Planck Society. To improve the matching, Transformative
Agreement Data were automatically enriched with ROR-IDs from associated
organisations according to OpenAlex's institution entity data.

\begin{figure}[ht!]

{\centering \includegraphics[width=0.99\linewidth,]{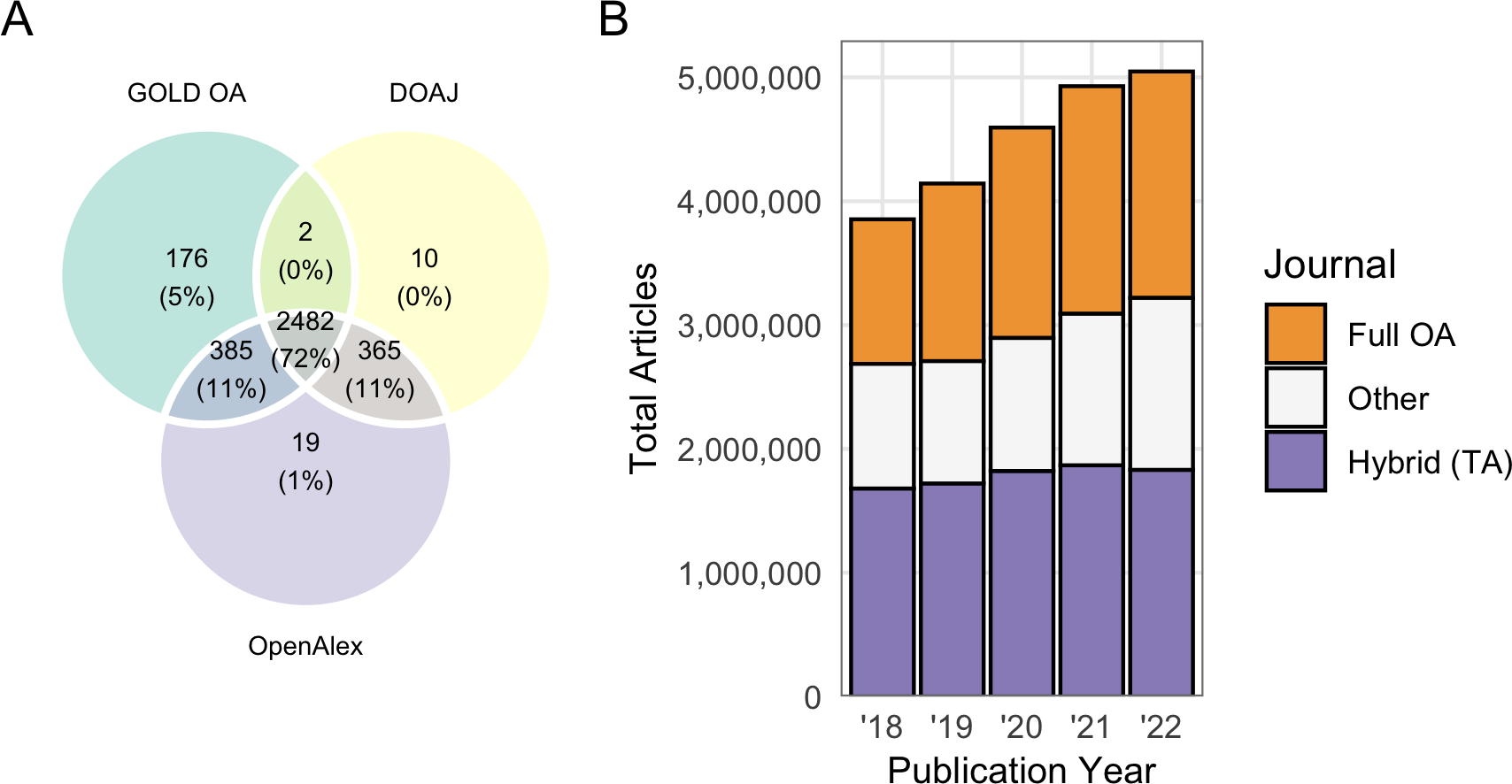} 

}

\caption{Initial data characteristics. (A) Full open access journals included in transformative agreements by evidence source Directory of Open Access Journals (DOAJ), OpenAlex and the Bielefeld GOLD OA list. (B) Number of articles in Crossref by journal types. The purple bars show the overall article volume of hybrid journals in transformative agreements, which were initially included in the study, in comparision with full open access journals according to OpenAlex. The remainder represents articles in subscription-based journals not covered by transformative agreements.}\label{fig:method_fig}
\end{figure}

The so compiled data set consists of 8,922,146 articles published in
12,857 hybrid journals included in at least one transformative agreement
between 2018 and 2022 (see purple bars in Figure \ref{fig:method_fig}B).
These hybrid journals in transformative agreements represented 40\% of
the total global output over the same time period, according to
Crossref, while fully open access journals recorded 35\%.

\subsection{Data analysis}\label{data-analysis}

Throughout this mostly automated data gathering and analysis process,
Tidyverse tools (Wickham et al., 2019) for the R programming language (R
Core Team, 2020) were used. The resulting data are openly available
through an R data package, hoaddata (Jahn, 2023). Following Marwick et
al. (2018), hoaddata contains not only the datasets used in the data
analysis. It also includes code used to compile the data by connecting
it to a cloud-based Google BigQuery data warehouse, where scholarly big
data from Crossref, OpenAlex and Unpaywall were made available, using
bigrquery (Wickham \& Bryan, 2023). To increase computational
reproducibility, data aggregation through hoaddata was automatically
carried out using GitHub Actions, a continuous integration service.

\section{Results}\label{results}

\subsection{Overview}\label{overview}

Between 2018 and 2022, 11,189 out of 12,857 hybrid journals in
transformative agreements published at least one open access article
under a Creative Commons license. These eligible 11,189 hybrid journals
constituted the foundation for the subsequent analyses. Collectively,
they provided open access to 742,369 out of 8,146,958 articles during
the investigated period, representing a five-year open access proportion
of 9.1\%. Authors who could make use of transformative agreements at the
time of publication contributed 328,957 (44\%) open access articles.
Overall, this investigation was able to establish a link between open
access articles and eligible institutions for 394 out of 502 (78\%)
active transformative agreements between 2018 and 2022 through the
linking of first author affiliations with agreement data.

\begin{figure}[ht!]

{\centering \includegraphics[width=0.99\linewidth,]{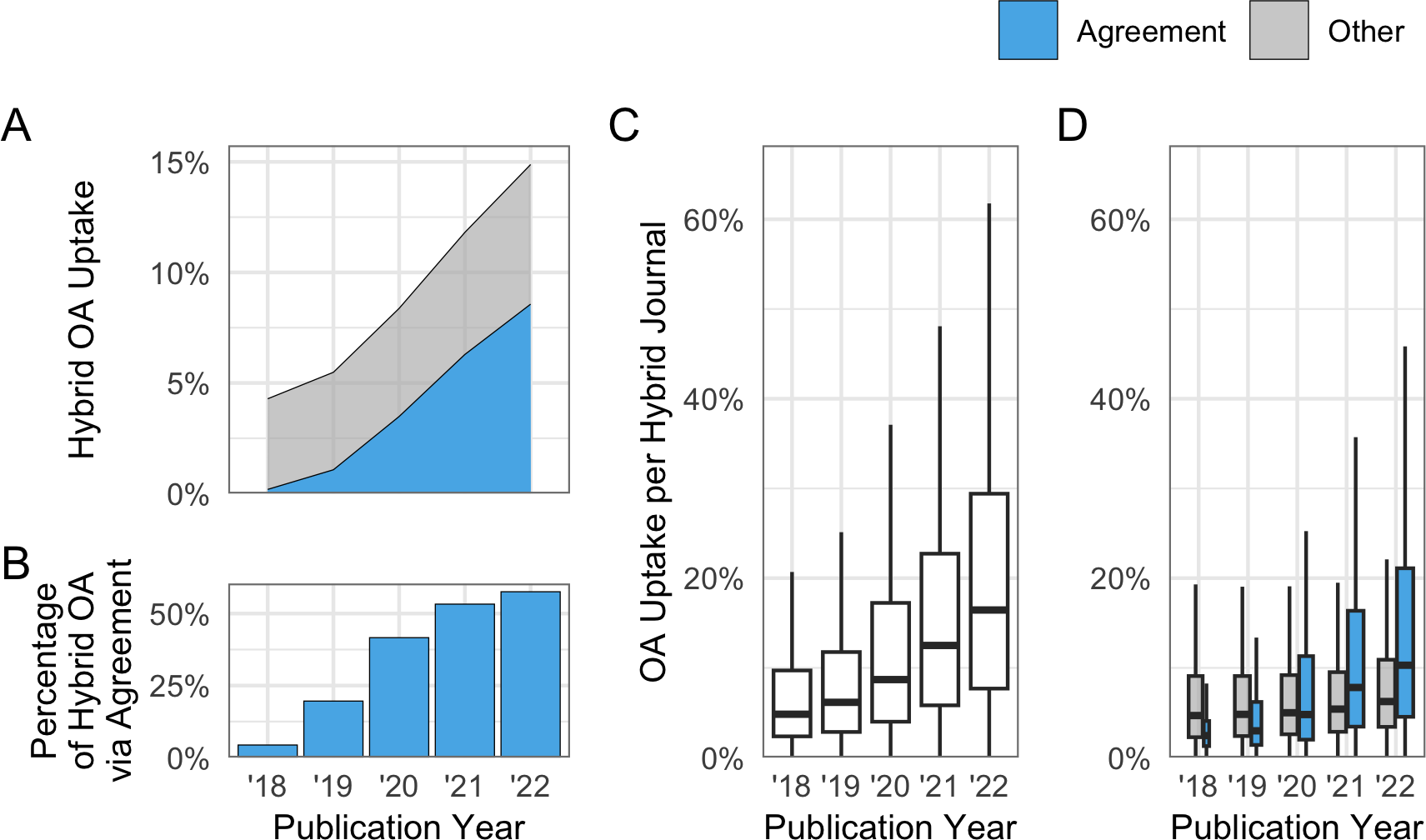} 

}

\caption{Relative growth of open access in hybrid journals in transformative agreements between 2018 and 2022 per publication year. The blue areas represent open access through transformative agreements, the grey areas depict open access articles where no link to an agreement could be established. (A) Proportion of open access articles in hybrid journals per year. (B) Percentage of hybrid open access via agreements per year. Boxplots show the proportion of open access articles by individual hybrid journals (C) and individual open access uptake rates by individual hybrid journals and open access funding (D) per publication year. The individual outliers are not shown. Note that data on transformative agreements ending before June 2021 were not available for this study.}\label{fig:results_overview}
\end{figure}

Figure \ref{fig:results_overview}A shows a moderate growth in the
proportion of open access articles in hybrid journals, comparing the
overall open access uptake and the impact of transformative agreements
on this trend. Over the five-years period from 2018 to 2022, open access
in hybrid journals increased from 4.3\% (n = 65,486) to 15\% (n =
249,511). Simultaneously, the total article volume of the investigated
journals rose from 1,528,051 in 2018 to 1,676,928 in 2022.

Figure \ref{fig:results_overview}B highlights that the majority of
hybrid open access was made available through transformative agreements
in 2021 and 2022. In 2022, 143,615 out of 249,511 open access articles
were from eligible authors, representing 58\%. However, there was also a
notable increase in open access provision through other means,
presumably publication fees being not invoiced through transformative
agreements as indicated by the grey area in \ref{fig:results_overview}A,
from 4.1\% (n = 62,625) in 2018 to 6.3\% (n = 105,896) in 2022.

Figure \ref{fig:results_overview}C depicts the substantial variations
among the hybrid journals included in transformative agreements in terms
of open access uptake. Although the median generally follows the trend
shown in Figure \ref{fig:results_overview}A, the farther stretch of
upper quartiles and whiskers over the years illustrates that an
increasing number of journals published an above-average proportion of
open access articles. In 2022, 25\% of hybrid journals (n = 2,576) had
an open access uptake of 29\%, and 6.6\% of journals (n = 744) provided
the majority of their articles under a CC license in the same year. On
average, these journals were smaller (M = 75, SD = 186) than those with
an open access share below 50\% (M = 164, SD = 347).

When comparing the impact of open access through transformative
agreements across journals, it shows that for many journals, these
agreements substantially contributed to the growth of open access over
the years (Figure \ref{fig:results_overview}D). Despite the rise in
transformative agreements, it is worth noting that other means of
publishing open access remained common across the investigated hybrid
journals. In total, 9,153 journals published open access articles from
authors affiliated with institutions without transformative agreements
in place, while 8,780 journals published at least one open access
article through a transformative agreement in the same year.

\subsection{Publishing market}\label{publishing-market}

Analysing hybrid open access across publishers between 2018 and 2022
reveals a large market concentration. Although 48 publishers offered
transformative agreements, the three commercial publishers Elsevier,
Springer Nature, and Wiley accounted for 49\% of hybrid journals,
representing 5,144,308 or 63\% of the total article volume (see Table
\ref{tab:publisher_league_table}). Together, they published 500,878 or
66\% of the open access articles in hybrid journals. Elsevier, Springer
Nature, and Wiley made 243,891 articles open access in hybrid journals
through transformative agreements, resulting in an even larger market
share of 74\%.

\begin{table}[H]

\caption{\label{tab:publisher_league_table}Hybrid open access through transformative agreements market shares 2018-2022}
\centering
\begin{tabular}[t]{lrlrlrlrl}
\toprule
\multicolumn{1}{c}{ } & \multicolumn{2}{c}{Hybrid journals} & \multicolumn{2}{c}{Articles} & \multicolumn{2}{c}{OA Articles} & \multicolumn{2}{c}{TA OA Articles} \\
\cmidrule(l{3pt}r{3pt}){2-3} \cmidrule(l{3pt}r{3pt}){4-5} \cmidrule(l{3pt}r{3pt}){6-7} \cmidrule(l{3pt}r{3pt}){8-9}
Publisher & Total & \% & Total & \% & Total & \% & Total & \%\\
\midrule
Elsevier & 1,936 & 17 & 2,770,826 & 33.8 & 172,723 & 22.9 & 60,440 & 18.3\\
Springer Nature & 2,274 & 20 & 1,330,430 & 16.2 & 175,432 & 23.3 & 100,008 & 30.3\\
Wiley & 1,410 & 12.4 & 1,043,052 & 12.7 & 152,723 & 20.3 & 83,443 & 25.3\\
Other & 5,767 & 50.6 & 3,061,337 & 37.3 & 252,523 & 33.5 & 86,294 & 26.1\\
\bottomrule
\end{tabular}
\end{table}

However, there are notable differences between the three large
publishers. Although Elsevier published the largest volume of articles
(n = 2,770,826; 34\%), it recorded a comparably low number of open
access articles, including those that are associated with transformative
agreements. In contrast, Springer Nature and Wiley provided open access
to a larger proportion of their articles (13\% of Springer Nature
articles and 15\% of Wiley articles were open access), leading to higher
open access market shares (23\% Springer Nature resp. 23\% Wiley). This
difference between Elsevier on the one hand and Springer Nature and
Wiley on the other can be attributed to transformative agreements, as
the latter made the majority of their open access articles available
through such deals (57\% Springer Nature resp. 55\% Wiley).

\begin{figure}[ht!]

{\centering \includegraphics[width=0.99\linewidth,]{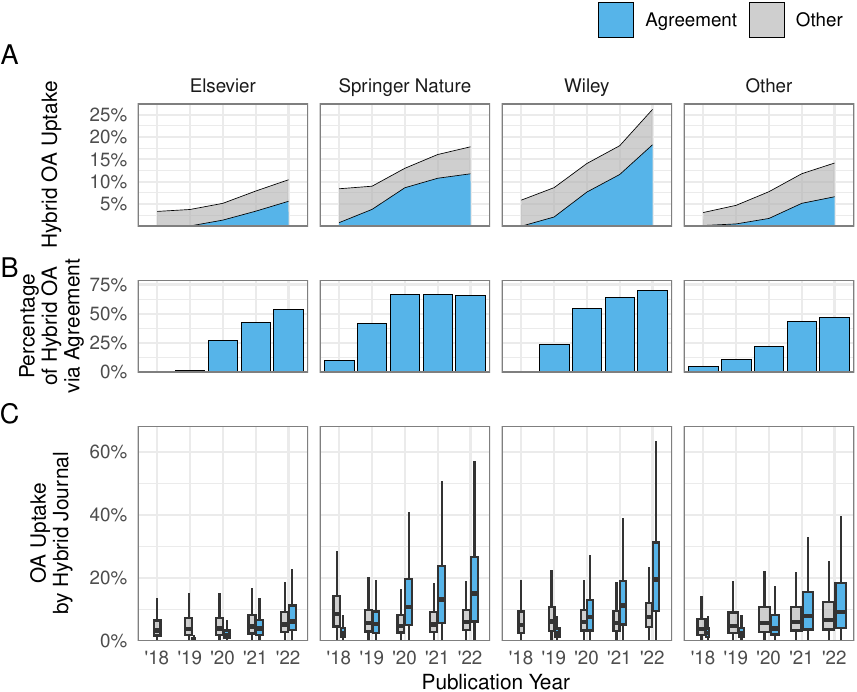} 

}

\caption{Development of open access in hybrid journals included in transformative agreements between 2018 and 2022 by publishers. The blue areas represent open access through transformative agreements, the grey areas depict open access articles where no link to an agreement could be established. (A) Proportion of open access articles in hybrid journals per year and publisher. (B) Percentage of hybrid open access via agreements per year and publisher. Boxplots (C) show individual open access uptake rates by individual hybrid journals and open access funding per publication year and publisher. The individual outliers are not shown. Note that data on transformative agreements ending before June 2021 were not available for this study.}\label{fig:publisher_figure}
\end{figure}

Figure \ref{fig:publisher_figure} takes a closer look into the growth of
hybrid open access across publishers by year, with a focus on open
access enabled by transformative agreements. Although all publishers
show a general long-term trend towards transformative agreements, Figure
\ref{fig:publisher_figure}A and B indicate that Wiley experienced a
substantial increase in its open access share from 5.9\% (n = 11,628) in
2018 to 26\% (n = 53,503) in 2022. In contrast, Elsevier's hybrid
journals demonstrated a more modest increase, from 3.3\% (n = 16,872) in
2018 to 10\% (n = 60,821) in 2022, which is a relatively low open access
share compared to the general trend. In 2018, Springer Nature had the
largest open access proportion among the three publishers of 8.4\% (n =
19,701), but experienced a relatively slower growth, resulting in 18\%
(n = 52,616) of articles being open access in Springer Nature hybrid
journals in 2022.

The varying degrees of adoption of open access across the three major
publishers can be attributed to distinct approaches to transformative
agreements. Springer Nature, for example, started offering open access
agreements for its hybrid journal portfolio to selected consortia such
as the Max Planck Society, the Swedish Bibsam consortium and the Finnish
FinELib consortium in 2015 under the name Springer Compact\footnote{\url{https://web.archive.org/web/20180414062853id_/http://www.liber2015.org.uk/wp-content/uploads/2015/03/Springer-Compact.pdf}}.
Prior to this, Springer had some pilot agreements with a small number of
institutions, including the University of Göttingen (Schmidt \& Shearer,
2012). However, the Springer Compact agreements were not included in the
data as they concluded before the start of the transformative agreement
data collection in June 2021. Nonetheless, the results demonstrate the
importance of agreements for Springer Nature's hybrid open access
business over the five-years period (Figure
\ref{fig:publisher_figure}B). In 2022, 66\% (n = 34,725) of open access
in Springer Nature hybrid journals was enabled through transformative
agreements. In the same year, 70\% (n = 37,316) of Wiley's open access
articles could be linked to transformative agreements. By contrast,
Elsevier published a comparatively lower proportion of its open access
articles through transformative agreements in 2022 (n = 32,627; 54\%).

The increasing trend towards transformative agreements can also be
observed at the journal-level (Figure \ref{fig:publisher_figure}C).
While no substantial differences between open access enabled through
transformative agreements and other revenue sources could be seen across
Elsevier's portfolio, the distribution of open access across Springer
Nature and Wiley hybrid journals indicates that the growth is not
limited to a few journals but extends across the portfolio. In
particular, Wiley's upper quantile, which represents the top 25\% of
journals in terms of the proportion of open access articles from
transformative agreements, increased markedly from 13\% in 2020 to 31\%
in 2022. Simultaneously, the median proportion increased from 7.5\% to
19\%. It is interesting to note that a small but increasing number of
journals from these two publishers provide open access to the majority
of articles through transformative agreements. Wiley recorded 68 and
Springer Nature 102 hybrid journals with an open access share above
50\%, which could be attributed solely to transformative agreements.
Upon inspection, these journals were mainly society or local language
journals with small annual article volumes.

\subsection{Journal subjects}\label{journal-subjects}

Table \ref{tab:subject_summary_table} presents a high-level overview of
hybrid open access by ASJC subject area using fractional counting to
account for journals belonging to more than one category. Between 2018
and 2022, most hybrid journals with at least one open access article
could be attributed to the Social Sciences category, which also includes
Arts and Humanities. However, these journals published the fewest number
of articles, whereas Physical Sciences journals recorded the most
articles, followed by Health Sciences and Life Sciences. In terms of
open access, Physical Sciences journals accounted for more than
one-third of the articles published in the five-year period, followed by
Health Sciences, Social Sciences and Life Sciences.

\begin{table}[H]

\caption{\label{tab:subject_summary_table}Hybrid open access through transformative agreements by journal subject 2018-2022}
\centering
\begin{tabular}[t]{lrlrlrlrl}
\toprule
\multicolumn{1}{c}{ } & \multicolumn{2}{c}{Hybrid journals} & \multicolumn{2}{c}{Articles} & \multicolumn{2}{c}{OA Articles} & \multicolumn{2}{c}{TA OA Articles} \\
\cmidrule(l{3pt}r{3pt}){2-3} \cmidrule(l{3pt}r{3pt}){4-5} \cmidrule(l{3pt}r{3pt}){6-7} \cmidrule(l{3pt}r{3pt}){8-9}
Journal subject & Total & \% & Total & \% & Total & \% & Total & \%\\
\midrule
Health Sciences & 2,342 & 22.5 & 1,998,045 & 28 & 199,265 & 27.7 & 81,913 & 25.7\\
Life Sciences & 1,399 & 13.4 & 1,080,346 & 15.1 & 133,526 & 18.6 & 48,570 & 15.2\\
Physical Sciences & 2,693 & 25.9 & 3,111,711 & 43.6 & 247,515 & 34.4 & 110,933 & 34.8\\
Social Sciences & 3,967 & 38.1 & 953,084 & 13.3 & 138,388 & 19.3 & 77,496 & 24.3\\
\bottomrule
\end{tabular}
\end{table}

\begin{figure}[ht!]

{\centering \includegraphics[width=0.99\linewidth,]{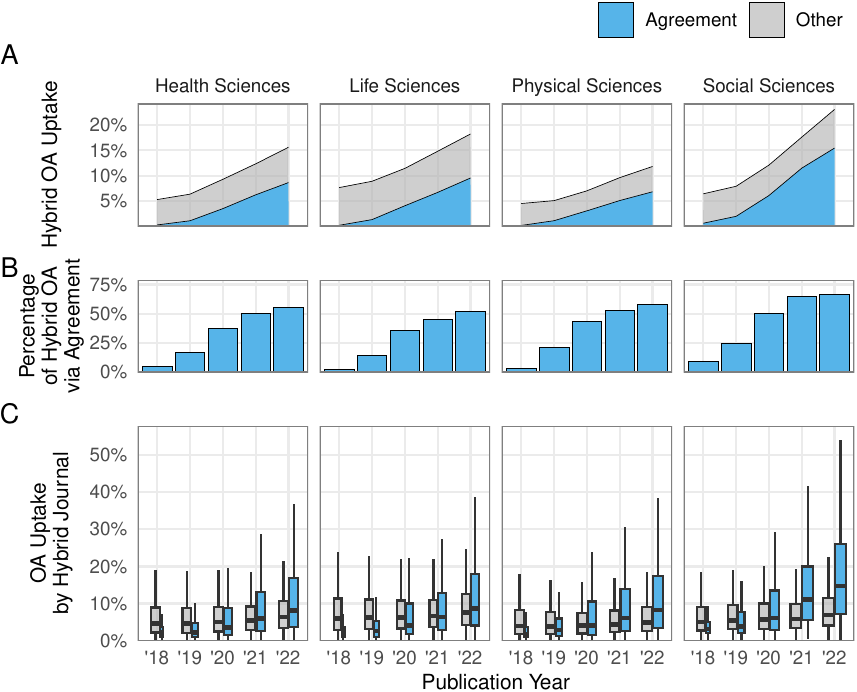} 

}

\caption{Development of open access in hybrid journals in transformative agreements between 2018 and 2022 by ASJC subject area. The blue areas represent open access through transformative agreements, the grey areas depict open access articles where no link to an agreement could be established. (A) Proportion of open access articles in hybrid journals per year and subject area. (B) Percentage of hybrid open access via agreements per year and subject area. (C) Boxplots show individual open access uptake rates by individual hybrid journals and open access funding per publication year and subject area. The individual outliers are not shown. Note that data on transformative agreements ending before June 2021 were not available for this study.}\label{fig:subject_panel}
\end{figure}

Figure \ref{fig:subject_panel} presents the relative growth of hybrid
open access by subject area between 2018 and 2022. In particular, Social
Sciences including Arts and Humanities journals accounted for the
strongest growth in the five-years period from 6.4\% (n = 8,361) to 23\%
(n = 51,938), followed by the Life Sciences from 7.6\% (n = 15,003) to
18\% (n = 39,494) , Health Sciences from 5.3\% (n = 18,279) to 16\% (n =
63,089) and Physical Sciences from 4.5\% (n = 22,364) to 12\% (n =
85,428). Growth in Social Sciences category can be largely attributed to
transformative agreements. In 2022, two-thirds of open access articles
(67\%, n = 34,759) were published by the first authors affiliated with
participating institutions (see \ref{fig:subject_panel}B).

Figure \ref{fig:subject_panel}C shows that this trend was consistent
across hybrid journals belonging to the ASJC Social Sciences category.
In 2022, 25\% of Social Sciences journals provided open access to at
least every fourth article exclusively through transformative
agreements. However, hybrid open access through transformative
agreements played a comparably lesser role in Life Sciences and Health
Sciences. In these two subject areas, only about half of the open access
articles can be linked to these agreements, both overall and on a median
average across journals. In contrast, the majority of Physical Science
journals show an increase of open access through transformative
agreements compared to other options to publish open access in hybrid
journals.

\subsection{Comparing countries}\label{comparing-countries}

Between 2018 and 2022, high-income countries almost exclusively
dominated hybrid open access publishing through transformative
agreements. To discern socio-economic differences, these countries were
grouped according to their membership in the Organisation for Economic
Co-operation and Development (OECD). Overall, first authors affiliated
with institutions from OECD member countries published 602,050 open
access articles in hybrid journals, representing 81\% of the
investigated open access articles. This disparity between OECD nations
and other countries becomes even more evident when considering open
access through transformative agreements, as 310,712 of 328,957, or 94\%
of open access articles were associated with such agreements.

\begin{figure}[ht!]

{\centering \includegraphics[width=0.99\linewidth,]{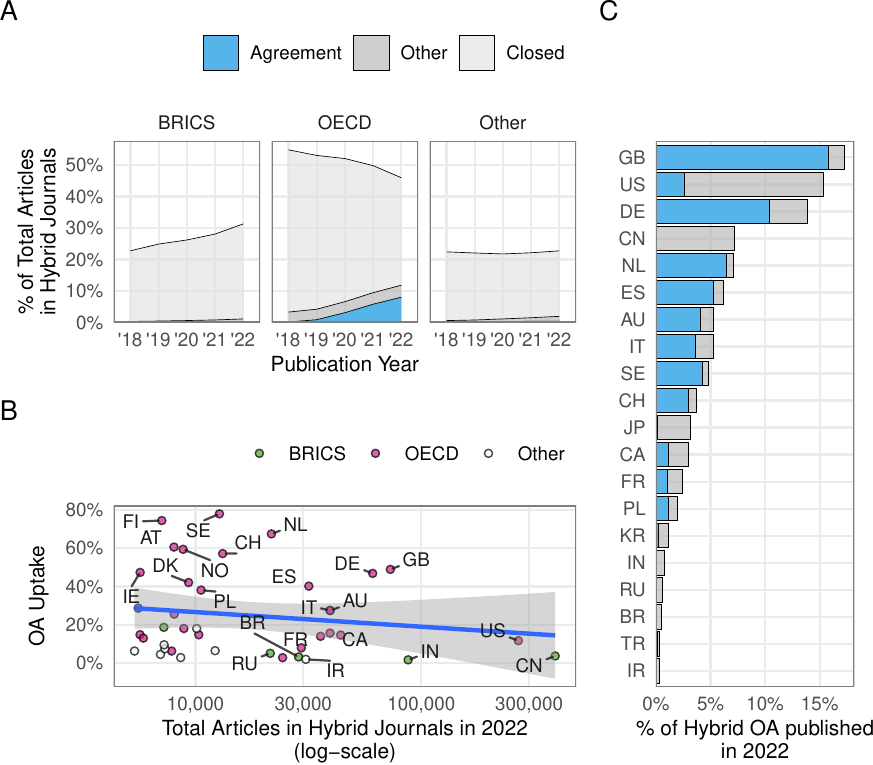} 

}

\caption{Development of hybrid open access publishing by country. (A) presents the relative number of articles published in hybrid journals included in transformative agreement by year, distinguishing between BRICS as of 2022, OECD and other countries. (B) Scatterplot contrasts total articles with open access article volume in 2022, by country and its OECD (purple-colored points) or BRICS (green) membership. Trend line obtained from linear regression, grey area show pointwise symmetric 95\% confidence bands. (C) Hybrid open access market share in 2022 by country. In (A) and (C), the blue areas represent open access through transformative agreements, the grey areas depict open access articles where no link to an agreement could be established. The remainder shows closed access articles. Country names are represented as ISO two-letter country codes.}\label{fig:country_patch}
\end{figure}

Figure \ref{fig:country_patch}A shows the development of hybrid open
access publishing by country, comparing the OECD area with the BRICS, an
intergovernmental organisation, which comprised Brazil, Russia, India,
China, and South Africa as of 2022. The residual category ``Other''
includes the remaining countries. Full counting was applied to account
for multiple country affiliations (Hottenrott et al., 2021). Open access
in hybrid journals increased in particular because of publications from
first authors affiliated with institutions from the OECD area, from
52,154 in 2018 to 202,787 in 2022. This growth was largely driven by
transformative agreements. Their share rose from 5.5\% in 2018 (n =
2,859) to 68\% (n = 137,815) in 2022. In contrast, BRICS recorded a low
uptake, moderately growing from 1.6\% in 2018 to 3.7\% in 2022.

Despite the rise of open access across OECD countries, the overall
publication output decreased sharply, dropping to 786,903 in 2022 after
peaking 892,197 articles in 2020, which saw a massive growth in
literature related to COVID-19, driven largely by researchers in the
United States surpassing China and other countries (Ioannidis et al.,
2021). In stark contrast, the number of articles published in hybrid
journals by first authors affiliated with institutions from BRICS
countries increased steadily over the years, from 356,632 in 2018 to
535,828 in 2022. This resulted in an increase in market share of the
BRICS area from 23\% to 31\% between 2018 and 2022, whereas that of the
OECD area decreased from 55\% to 46\% during the same period. Upon
closer examination, this trend can be observed across all of the three
largest publishers, although the shift towards BRICS is particularly
evident in Elsevier's hybrid journal portfolio, particularly with regard
to articles published in Physical Sciences journals. While OECD
publication output in Elsevier's Physical Sciences journals declined
from 112,822 articles in 2018 to 103,766 in 2022, BRICS output increased
from 104,654 to 171,713 in the same five-year period. Furthermore, OECD
publication output in Health Sciences and Life Sciences journals
stagnated after peaking in 2020, which saw an increase in publications
due to the impact of the global COVID-19 pandemic, in particular from
the United States (Ioannidis et al., 2021).

To illustrate the situation in 2022, Figure \ref{fig:country_patch}B
compares the total publication output with the number of open access
articles. With 391,530 articles, China was the most productive country,
followed by the United States (n = 268,965) and India (n = 87,428). In
contrast, Western and Northern European countries published a
considerably high number of open access articles. Particularly, Nordic
countries, the Netherlands and Austria recorded above-average open
access shares, as indicated by the linear trend line. As shown in Figure
\ref{fig:country_patch}C, transformative agreements contributed to these
market positions. Interestingly, the United States had a notable open
access market share of 15\%, although transformative agreements
contributed to a lesser extent. Similarly, China's open access market
share of 7.2\% in 2022 was comparable to that of the Netherlands, which
was 7.1\%.

\begin{figure}[ht!]

{\centering \includegraphics[width=0.99\linewidth,]{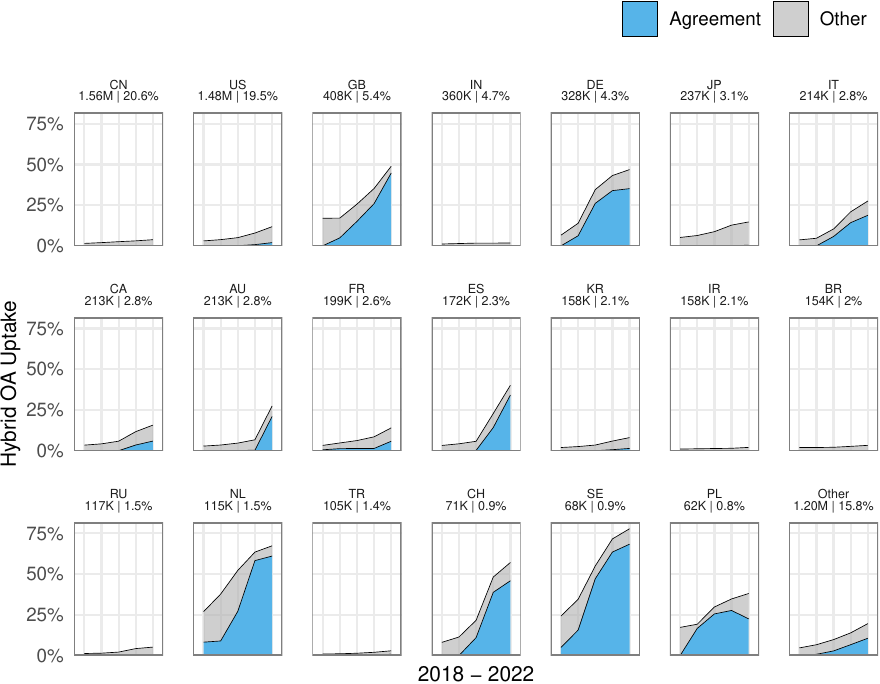} 

}

\caption{Development of open access in hybrid journals in transformative agreements between 2018 and 2022, by the Top 20 most productive countries in terms of total articles published in the five-years period. Blue areas represent open access through transformative agreements, the grey areas depict open access articles where no link to an agreement could be established. Country names are represented as ISO two-letter country codes. Facet subheadings show the total number of articles and corresponding market share.}\label{fig:country_top_20_plot}
\end{figure}

Figure \ref{fig:country_top_20_plot} illustrates the development of
hybrid open access from 2018 to 2022, highlighting the top 20 most
productive countries in terms of articles published in hybrid journals
that were included in transformative agreements over the five-year
period. Notably, the Netherlands (27\%), Sweden (24\%), Poland (17\%)
and Great Britain (17\%) exhibited a relatively high level of uptake in
2018 which continued to increase in the following years. In 2022, Sweden
had the highest proportion of open access relative to its publication
output (78\%), followed by the Netherlands (67\%) and Switzerland
(57\%), with these countries benefiting from transformative agreements.
In Germany, however, hybrid open access only began to increase from 2019
onwards after the successful negotiation of nationwide agreements with
Wiley (July 2019) and Springer Nature (January 2020). Prior to this,
only a few organisations had agreements in place, for example, the Max
Planck Society with Springer Compact.

Since 2021, there has been a general trend towards hybrid open access
among many high-income countries, driven primarily by transformative
agreements. However, the proliferation of transformative agreements
differed across these countries. Additionally, publication limits or
eligibility criteria for institutions and article types may explain why
even countries with widespread agreement implementation such as Sweden
or the Netherlands did not achieve 100\% hybrid open access.
Interestingly, in Japan and the United States, options other than
transformative agreements were the main drivers for the increase in
hybrid open access. Once again, the graph highlights countries with low
hybrid open access, particularly non-OECD countries, where only a few or
no agreements were in place.

\section{Discussion}\label{discussion}

The primary aim of this study was to assess the uptake of open access in
hybrid journals included in transformative agreements, which were
introduced as a temporal means to support the transition of
subscription-based academic publishing to full open access. This study
presents a novel approach based on open data, which leverages metadata
on over 700 agreements and nine million journal articles to estimate the
extent to which transformation agreements contribute to the transition
of this journal business model. The results highlight a strong growth in
open access between 2018 and 2022, driven by an increasing number
transformative agreements. However, the majority of research literature
published in hybrid journals in this five-year period remained behind
publisher paywalls. Growth in the adoption of open access in hybrid
journals, in particular through transformative agreements, can be
largely attributed to three large commercial publishers -- Elsevier,
Springer Nature, and Wiley -- but varies substantially across journals,
publishers, disciplines, and country affiliations. Despite the
limitations of the data, the findings indicate that the current level of
implementation of transformative agreements is insufficient to bring
about a large-scale transition to full open access.

A key finding of this analysis is that transformative agreements
maintain market concentration. Specifically, the three largest
commercial publishers Elsevier, Springer Nature, and Wiley dominate,
particularly with regard to open access provided through transformative
agreements. Together, the three publishers accounted for three-fourths
of open access articles through transformative agreements, while
recording about half of active hybrid journals included in
transformative agreements between 2018 and 2022. This observation aligns
with previous research on shifts in the publishing market following the
introduction of funding opportunities for hybrid open access (Butler et
al., 2023; Jahn \& Tullney, 2016; Shu \& Larivière, 2023). Additionally,
the results confirm previously observed variations by publisher, with
Elsevier exhibiting a different development than Springer Nature and
Wiley (Butler et al., 2023; Jahn et al., 2022). However, it must be
noted that the focus of transformative agreements on publishers with
large hybrid journal portfolios is intentional (Campbell et al., 2022).
Because of transformative agreements with a few large publishers,
national consortia were able to substantially increase their country's
annual open access article volume (Bosman, Jonge, et al., 2021; Brayman
et al., 2024; Huang et al., 2020; Pinhasi et al., 2021; Taubert et al.,
2023; Wenaas, 2022; Widding, 2024).

Moreover, this study presents varying levels of open access uptake
through transformative agreements across journals, which can be
attributed to the alignment of authors' affiliations and the
availability of such agreements at their institutions. In line with
previous research findings (Butler et al., 2023; Jahn et al., 2022;
Wenaas, 2022), high uptake rates were observed across Social Sciences
and Humanities hybrid journals. However, it is important to emphasise
that these hybrid journals do not encompass the entire field. For
example, Khanna et al. (2022) found that approximately 60\% of journals
utilising the open-source publishing platform Open Journal Systems (OJS)
fall within the Social Sciences and Humanities. In these fields, the
majority of full open access journals are so-called ``Diamond OA
journals'', which do not charge publication fees (Bosman, Frantsvåg, et
al., 2021). Rather, this result can be more accurately attributed to the
substantial proportion of authors from high-income countries who publish
in these hybrid journals included in transformative agreements,
particularly when journals are of local relevance, such as by belonging
to national societies or regional research scope.

Surprisingly, the total publication output of high-income countries
belonging to the OECD in hybrid journals declined substantially after
peaking in 2020, whereas that of BRICS countries continuously grew from
2018 to 2022. Because the BRICS expansion can be exclusively attributed
to closed access articles, this development has the potential to hinder
the transition of academic journal publishing to full open access
through transformative agreements and demands discussion. In China, the
country with the highest volume of articles in hybrid journals, only
limited research funding to pay for open access is available, with
expenditures for open access publishing surpassing subscription costs at
some universities (Shu \& Larivière, 2023). Furthermore, the focus of
Chinese authors on established journals may contribute to this trend
(Zhang et al., 2022). Although China supports Plan S (Schiermeier,
2018), this is not reflected in the data in terms of open access uptake
in hybrid journals. Market observers also do not expect a broad
implementation of transformative agreements in China in near future
(Owens, 2024). As highlighted by Koley \& Lala (2023), India faces
similar challenges in terms of the availability of resources to pay for
open access. At the same time, access to research literature is a
pressing issue, which is addressed by the ``Indian one nation, one
subscription'' policy proposal. However, this policy focuses on
centrally negotiated subscriptions and does not entail open access.

But open access uptake also differs among OECD countries. In the United
States, for example, hybrid open access, including transformative
agreements, plays a lesser role than in some European countries. Between
2017 and 2021, hybrid open access contributed the least to openly
available federally funded research articles (Schares, 2023). While some
university consortia, such as the California Digital Library, have
signed transformative agreements, others have attempted to depart from
big deals and unbundle large journal portfolios to address cost
increases (Brainard, 2021; Schares, 2022). Despite the relatively low
penetration of transformative agreements in China and the United States
compared to Europe, 22\% of open access in hybrid journals in 2022
originates from these two countries, indicating the availability of
funding sources for publication fees.

This large-scale study provides the first empirical evidence of the
influence of transformative agreements on the transition of hybrid
journals to full open access. However, several limitations need to be
acknowledged. From a data perspective, estimations of open access
through transformative agreements were established by linking author
affiliations with publicly available agreement data from cOAlition S,
and not through invoicing data, which is usually not shared. Because
data on corresponding authors, who are typically responsible for
facilitating open access publication, were not fully supported by
OpenAlex at the time of this study, first author affiliations were used
instead. This study is also unable to account for the various types of
transformative agreements due to a lack of data, especially regarding
article types and caps that limit the number of open access articles
covered. Furthermore, assessing the quality of the OpenAlex and Crossref
data used, particularly in terms of affiliations, corresponding
authorship, and article types, by combining them with established
bibliometric databases such as Scopus and Web of Science was beyond the
scope of this analysis (Chinchilla-Rodríguez et al., 2024; Visser et
al., 2021). It must therefore be emphasised that, unlike evaluations
from national consortia that could make use of invoices (Brayman et al.,
2024), this global overview can only provide an estimate of the support
for open access through transformative agreements. Despite these
limitations, the methodology was designed to underestimate, rather than
overestimate, the adoption of open access through transformative
agreements. To promote transparency, the data used in this study, along
with the code used for this analysis, are openly available.

Additionally, it must be noted that the study period was significantly
impacted by the COVID-19 pandemic, which led to an unprecedented number
of publications and a reduction in international collaboration
(Aviv-Reuven \& Rosenfeld, 2021), which could explain the observed
contrasting developments in OECD and BRICS countries. However, even
before the pandemic, growth in publications in Europe was only due to
internationally co-authored journal articles (Kwiek, 2021). Likewise,
inflows from China to the United States and European countries already
declined by 2020 (Zhao et al., 2023). Furthermore, the study design did
not consider emerging publication practices such as preprints (Fraser et
al., 2021) and special issues (Hanson et al., 2023), which have grown
rapidly since 2020. Lastly, it should be emphasised that the study did
not address financial shifts between subscriptions spending and open
access payments while analysing hybrid open access through
transformative agreements due to a lack of data on expenditures.

This study allows for multiple strands of further research. One is to
complement this large-scale study with more specific evidence from
individual countries or subjects, particularly those with low hybrid
open access rates. Incorporating full open access and subscription-based
journals, as well as considering global trends in scholarly migration
and collaboration could also be promising. Financial studies could build
on the study design and include subscription and open access expenditure
to assess the cost-effectiveness of transformative agreements, in
particular whether transformative agreements can integrate the
substantial amounts of individual payments for publication fees (Butler
et al., 2023; Wenaas, 2022), as well as potential changes in authors'
behaviour following the introduction of these agreements (Schmal, 2024).

This study has practical implications for research funding and
libraries. One concern should be the observed differences across
countries, particularly the relationship between socio-economic
development and open access adoption. Between 2018 and 2022, there was a
notable increase in the number of closed access articles published in
hybrid journals by authors from countries such as China and India, where
transformative agreements and open access funding options were not
widely available. In contrast, the introduction of country-wide
transformative agreements in numerous OECD countries led to a
substantial increase in open access, while their share of total articles
in hybrid journals decreased. This imbalance, whereby open access uptake
depends largely on a few countries, makes it less likely that collective
action towards a large-scale transition of hybrid journals to full open
access through transformation agreements will succeed within the next
years. Rather, and in accordance with recent British and Scandinavian
policy recommendations (Brayman et al., 2024; Widding, 2024), it
emphasises the importance of ongoing evaluation and adaptation of open
access strategies in view of global publishing trends and related
challenges to equity and journal quality (Rasmussen, 2023; Ross-Hellauer
et al., 2022).

From a data perspective, research institutions and funders need to push
towards price transparency in scholarly publishing to enable these
evaluations. The reporting of open access funding, including
transformative agreements, is also not harmonised, but often
crowd-sourced from various sources. To improve the assessment of
transformative agreements, libraries and publishers should collaborate
on standards and services to publicly share information about respective
journal portfolios, participating institutions and open access
invoicing, for example in the international context of ESAC and the
Barcelona Declaration on Open Research Information. Such collaboration
would help overcome the limitations of current approaches that derive
funding for open access from authorship data.

In summary, this study provides empirical insights into the development
of hybrid open access following the introduction of transformative
agreements. These results are important for both researchers and
stakeholders engaged in negotiating and evaluating these agreements. The
presented approach relies on open data, which enables follow-up studies
and open access monitoring activities to further explore the role of
transformative agreements in transitioning academic publishing to full
open access.

\section{Competing interests}\label{competing-interests}

The author declares no competing interests.

\section{Funding information}\label{funding-information}

This work was supported by the Deutsche Forschungsgemeinschaft (Grant
number 416115939).

\section{Data and code availability}\label{data-and-code-availability}

Source code analysis including data used is available on GitHub:
\url{https://github.com/njahn82/hoa_ta_effects}.

\section*{References}\label{references}
\addcontentsline{toc}{section}{References}

\phantomsection\label{refs}

\end{document}